\documentclass[%
reprint,
superscriptaddress,
nofootinbib,
 amsmath,amssymb,
 aps,
 prl,
 nolongbibliography,
]{revtex4-2}

\usepackage{subfiles}
\usepackage{graphicx}%
\usepackage{dcolumn}%
\usepackage{bm}%
\usepackage{soul}
\usepackage{bibunits}

\usepackage{subcaption}
\usepackage{xcolor}
\usepackage{lipsum}
\usepackage{hyperref}%
\hypersetup{
    colorlinks=true,
    linkcolor=blue,
    citecolor=blue,
    urlcolor=blue,
}

\makeatletter
\newcommand\footnoteref[1]{\protected@xdef\@thefnmark{\ref{#1}}\@footnotemark}
\makeatother

\newcommand*{\inlinesect}[1]{\emph{#1}.---}

\newcommand{\omegaAvg}{\langle\omega\rangle}
\newcommand{\meV}{\text{ meV}}

\newcommand{\keV}{\text{ keV}}
\newcommand{\MeV}{\text{ MeV}}
\newcommand{\GeV}{\text{ GeV}}

\newcommand{\MPa}{\text{ MPa}}
\newcommand{\fccHe}{$\beta$-helium}
\newcommand{\hcpHe}{$\alpha$-helium}
\newcommand{\omegamin}{\omega_\text{min}}
\newcommand{\csla}{\bar{c}_S^{LA}}

\begin{document}

\title{Pressure-Tunable Targets for Light Dark Matter Direct Detection: \\The Case of Solid Helium}

\author{Omar A. Ashour}          %
\affiliation{Department of Physics, University of California, Berkeley, California 94720, USA}
\affiliation{Materials Sciences Division, Lawrence Berkeley National Laboratory, Berkeley, CA 94720, USA}
\affiliation{Molecular Foundry, Lawrence Berkeley National Laboratory, Berkeley, CA, 94720, USA}

\author{Sin\'{e}ad M. Griffin}          %
\email{SGriffin@lbl.gov}
\affiliation{Materials Sciences Division, Lawrence Berkeley National Laboratory, Berkeley, CA 94720, USA}
\affiliation{Molecular Foundry, Lawrence Berkeley National Laboratory, Berkeley, CA, 94720, USA}

\date{\today}%

\begin{abstract}
We propose hydrostatic pressure---a well-established tool for tuning properties of condensed matter---as a novel route for optimizing targets for light dark matter direct detection, specifically via phonons. Pressure dramatically affects compressible solids by boosting the speed of sound and phonon frequencies. Focusing on helium---the most compressible solid---our \emph{ab initio} calculations illustrate how high pressure elevates helium from lacking single-phonon reach to rivaling leading candidates. Our work establishes pressure as an unexplored tuning knob for accessing lower dark matter mass regimes.
\end{abstract}

\maketitle

\inlinesect{Introduction}
Proposals for next-generation dark matter (DM) searches are extending into lower mass ranges, driven by advances in detector sensitivities that allow probing low-energy excitations in condensed matter systems \cite{kahnSearchesLightDark2022,mitridateSnowmassWhitePaper2023}. Traditional direct detection experiments based on nuclear recoils are ineffective in the sub-GeV DM mass range, as the kinetic energy deposition from the light DM to the heavier nuclei is minimal. This limitation has catalyzed research into low-energy excitations across various condensed matter systems including electronic transitions in semiconductors \cite{griffinExtendedCalculationDark2021,griffinSiliconCarbideDetectors2021,essigDirectDetectionSubGeV2016,hochbergAbsorptionLightDark2017,blancoDarkMatterDaily2021}, phonons in dielectrics \cite{griffinDirectionalDetectionLight2018,griffinMultichannelDirectDetection2020,trickleMultichannelDirectDetection2020,coskunerDirectionalDetectabilityDark2022,taufertshoferBroadRangeDirectionalDetection2023,campbell-deemDarkMatterDirect2022,mitridateEffectiveFieldTheory2024} and superfluids \cite{schutzDetectabilityLightDark2016,knapenLightDarkMatter2017,vonkrosigkDELightDirectSearch2023,youSignaturesDetectionProspects2023,hirschelSuperfluidHeliumUltralight2024}, as well as more exotic quantum materials such as superconductors \cite{hochbergDetectingSubGeVDark2019, hochbergDetectingUltralightBosonic2016, hochbergDirectionalDetectionLight2023, hochbergSuperconductingDetectorsSuperlight2016}, Dirac semimetals \cite{Hochberg2018,inzaniPredictionTunableSpinorbit2021,geilhufeDiracMaterialsSubMeV2020}, multiferroics \cite{roisingAxionmatterCouplingMultiferroics2021}, and topological axion insulators \cite{marshProposalDetectDark2019, schutte-engelAxionQuasiparticlesAxion2021}. For instance, phonons are the key target in the in-development TESSERACT collaboration for low-mass DM detection \cite{TESSERACT}.

Phonons are a promising avenue for light (sub-GeV) DM detection due to their sensitivity to various DM models, and their kinematic and energetic match with DM masses as low as a few keV through scattering \cite{griffinMultichannelDirectDetection2020} and a few meV through absorption \cite{mitridateEffectiveFieldTheory2024}. Notably, dark photon mediators couple to optical (gapped) phonons in polar semiconductors, such as Al$_2$O$_3$, with potential directional responses \cite{griffinDirectionalDetectionLight2018,coskunerDirectionalDetectabilityDark2022}. In contrast, hadrophilic scalar mediators are more strongly coupled to acoustic (gapless) phonons in both crystals and superfluids \cite{griffinMultichannelDirectDetection2020}. At low detector thresholds and small momentum transfer, single-phonon production is generally the dominant interaction in crystalline materials, whereas multiphonon processes dominate in superfluids, with considerable phase space suppression \cite{schutzDetectabilityLightDark2016,knapenLightDarkMatter2017}. However, single phonons usually have energies $\mathcal{O}(10-100)$ meV in crystals and thus require low sensing thresholds.

Hydrostatic pressure has long been exploited for exploring new phases in condensed matter and tuning properties of materials otherwise inaccessible at ambient conditions. By directly altering interatomic distances and electronic interactions, pressure can dramatically change the electronic, structural, and vibrational properties of matter. For instance, pressure is critical for accessing high-temperature superconductivity in hydrides \cite{bhattacharyyaImagingMeissnerEffect2024, chenHighTemperatureSuperconductingPhases2021, doluiFeasibleRouteHighTemperature2024, drozdovConventionalSuperconductivity2032015, drozdovSuperconductivity250Lanthanum2019, kongSuperconductivity243Yttriumhydrogen2021, maHighTemperatureSuperconductingPhase2022, troyanAnomalousHighTemperatureSuperconductivity2021} and for the discovering of topological phases in elemental lithium \cite{mackEmergenceTopologicalElectronic2019} and tellurium \cite{ideuePressureinducedTopologicalPhase2019}. Pressure is also used to control the performance of functional materials, with applications ranging from space-based photovoltaics \cite{iqbalAbinitioStudyPressure2022} to optoelectronic devices \cite{xuUsingPressureUnravel2023}.  

In this \emph{Letter}, we explore the use of hydrostatic pressure as a new tuning knob for controlling the reach of compressible solid targets as DM detectors. Pressure typically enhances the speed of sound and phonon frequencies in solids, with this effect being especially pronounced in highly compressible materials. We demonstrate this concept with solid helium-4, selected for its especially high compressibility and existing interest in superfluid helium as a DM target \cite{vonkrosigkDELightDirectSearch2023,youSignaturesDetectionProspects2023,hirschelSuperfluidHeliumUltralight2024}. At near-ambient pressures, solid helium-4 has limited single-phonon reach owing to its small phonon energies ($\lesssim$ 3 meV \cite{minkiewiczPhononSpectrumHcp1968}) and low speed of sound ($\sim 10^{-6}$ \cite{nosanowCalculationsSoundVelocities1965}). Our calculations show that by applying pressures up to 40 GPa, we can significantly enhance the speed of sound in helium by more than tenfold and phonon frequencies by nearly two orders of magnitude, dramatically improving its prospects for light DM detection. We perform density functional theory (DFT) calculations on two solid phases of helium-4, elucidating the effect of pressure on the structural and vibrational properties. We calculate the single-phonon reach and compare our results to the best-performing targets for a hadrophilic scalar mediator. Finally, we address the challenges and limitations of pressurized targets and examine how this unexplored degree of freedom could be best used in direct detection experiments.

\begin{figure*}
\includegraphics[width=\linewidth]{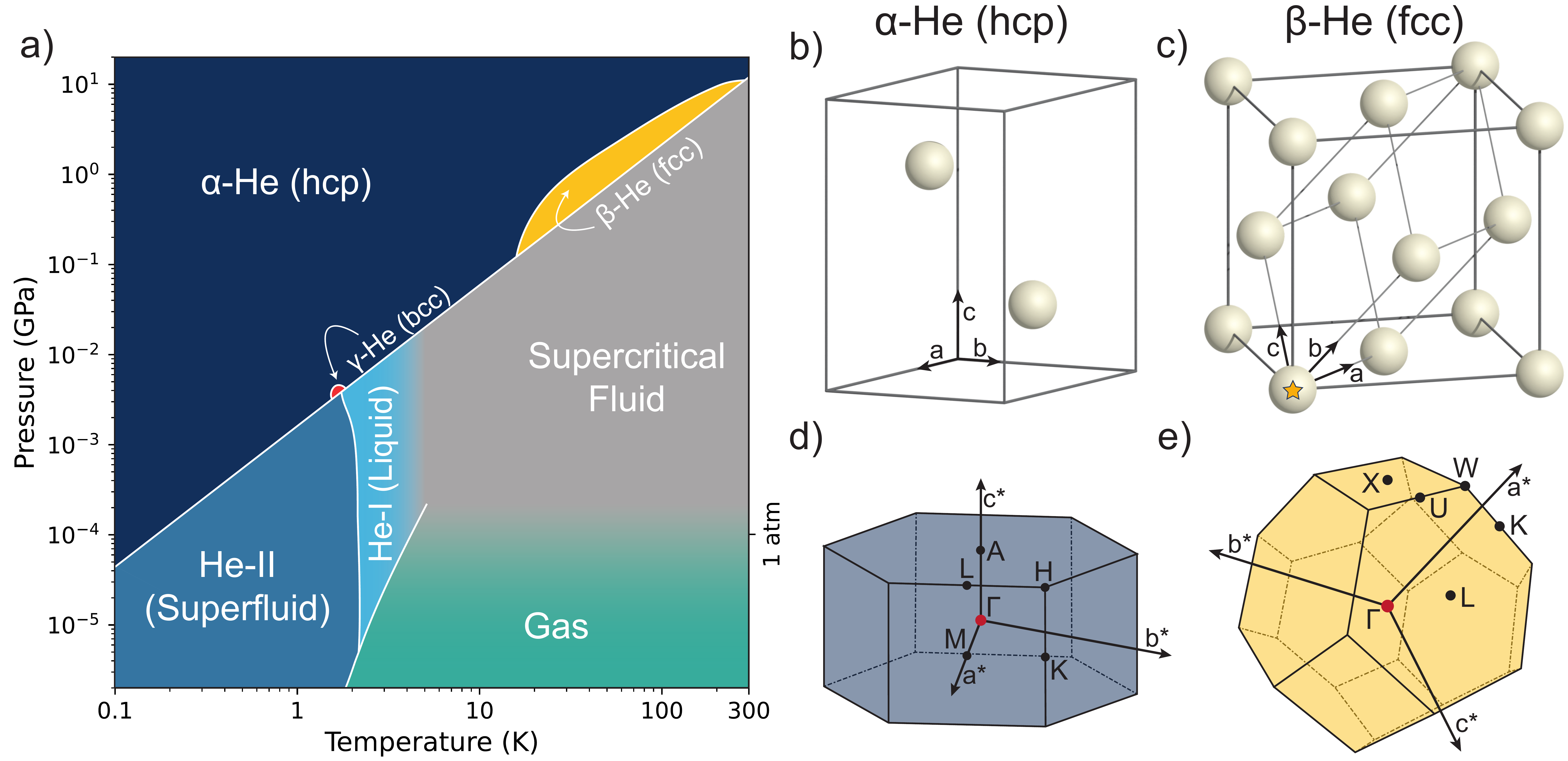}%
\caption{\label{fig:fig1} (a) Pressure-temperature phase diagram of helium-4, with experimental phase boundaries from Refs. \cite{loubeyreEquationStatePhase1993,vosHighpressureTriplePoint1990,osgoodInelasticNeutronScatteringBcc1972,sciverheliumCryogenics2015}, but we exaggerate the size of the bcc phase for visibility. (b, c) The unit cell of $\alpha$-helium and $\beta$-helium, respectively, with the primitive lattice vectors labeled by $a$, $b$, $c$. The primitive fcc cell contains one atom, marked by a $\star$. (d, e) The first Brillouin zones of the $\alpha$ and $\beta$ phases, respectively, with highlighted high symmetry points and reciprocal lattice basis vectors $a^*$, $b^*$, and $c^*$.}
\end{figure*}

\inlinesect{Modeling Solid Helium}
The pressure-temperature (PT) phase diagram of helium-4 is shown in Fig. \ref{fig:fig1}(a), indicating its three known solid phases. The hexagonally close-packed (hcp) $\alpha$ phase is the dominant solid phase, stable at pressures down to a few\MPa{} at sub-1 K temperatures, while the face-centered cubic (fcc) $\beta$ phase, is only stable at higher temperatures, $\mathcal{O}(10~\text{K})$, and much higher pressures, $\mathcal{O}(100 \MPa{})$ \cite{loubeyreEquationStatePhase1993}. Finally, the body-centered cubic (bcc) $\gamma$ phase is only stable in a minute sliver of the PT phase diagram \cite{vignosNewSolidPhase1961}, so we will focus exclusively on the $\alpha$ and $\beta$ phases. %

Studying the structural and vibrational properties of helium with DFT presents difficulties due to helium's significant zero-point motion and long-range interactions. The Born-Oppenheimer approximation used in DFT does not capture the former, and the latter is underestimated by standard semi-local exchange-correlation (XC) functionals. We tested 30 combinations of XC functionals and nonlocal corrections and verified our approach by comparing our DFT results to a variety of experimental measurements, including inelastic neutron scattering and Brillouin scattering. We find that the semi-local PBE functional \cite{PBE} combined with the D3M dispersion correction \cite{DFT-D3,DFT-D3M} provides excellent agreement for both phases across a large range of pressures, except for at low pressures which we do not include in this work (see the Supplemental Material (SM) \cite{supp} for details).

\begin{figure*}
\includegraphics[width=\linewidth]{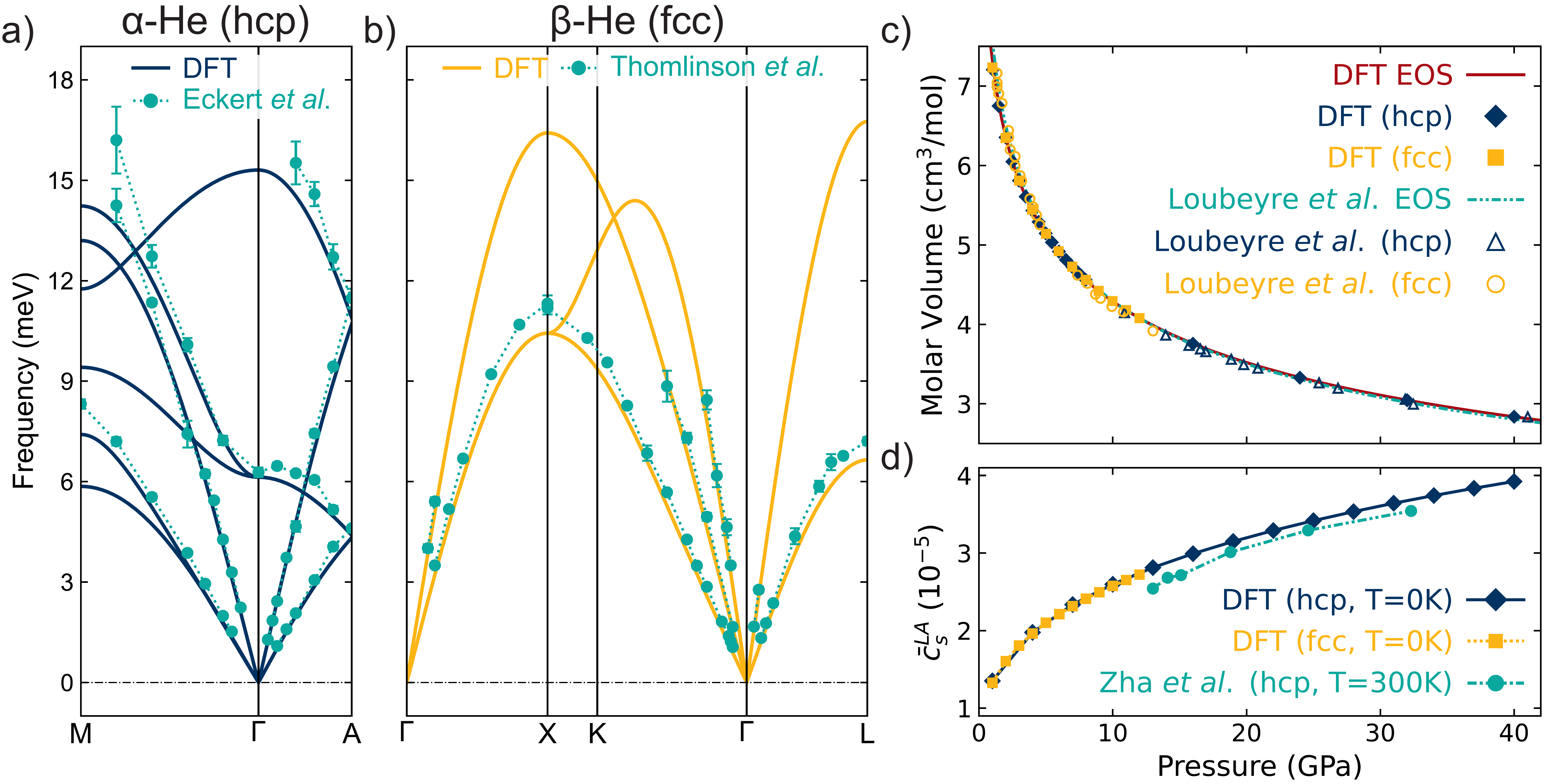}
\caption{\label{fig:fig2} (a,b) Phonon dispersion of \hcpHe{} (\fccHe{}) at a molar volume of 9.41 (9.03) cm$^3$/mol ($\sim$0.4-0.5 GPa). Navy (gold) lines are the DFT results for the $\alpha$ ($\beta$) phase, and teal circles represent experimental data at T = 10 K (22 K) from Ref. \cite{eckertLatticeDynamicsHcp1978} (\cite{thomlinsonInelasticNeutronScattering1978}), respectively. (c) DFT-calculated and experimental volumes at different pressures. DFT data is depicted by solid navy diamonds (gold squares) for the $\alpha$ ($\beta$) phase, and the experimental data \cite{loubeyreEquationStatePhase1993} by unfilled navy triangles (gold circles). The solid maroon (dashed teal) lines represent fits of the DFT (experimental) data to a Vinet equation of state \cite{loubeyreEquationStatePhase1993, vinetTemperatureEffectsUniversal1987,vinetUniversalEquationState1986a}. (d) The DFT-calculated speed of sound for the $\alpha$ (navy diamonds) and $\beta$ (gold squares) phases. Teal circles depict experimental data at 300K \cite{zhaElasticityDensehelium2004}.}
\end{figure*}

\inlinesect{Single-Phonon Scattering Rates}
We consider a benchmark model where real scalar DM $\chi$ couples to the proton $p$ and neutron $n$ via a real scalar mediator $\phi$ (neglecting any DM-electron coupling). The Lagrangian is given by 

\begin{equation}
\begin{aligned}
    \mathcal{L}= & \frac{1}{2}\left(\partial_\mu \phi\right)^2-\frac{1}{2} m_\phi^2 \phi^2+f_p \phi \bar{p} p+f_n \phi \bar{n} n \\
    & +\left(\frac{1}{2}\left(\partial_\mu \chi\right)^2-\frac{1}{2} m_\chi^2 \chi^2+\frac{1}{2} y_\chi m_\chi \phi \chi^2\right),
\end{aligned}
\vspace{0.01cm}
\end{equation}
where $m_\chi$ ($m_\phi$) is the mass of the DM (mediator), $f_p$ ($f_n$) is the coupling to the proton (neutron), and $y_\chi$ is the mediator-DM coupling. We set $f_p = f_n$ for the remainder of this work.

The dominant DM-matter interactions depend on the range of DM under consideration: for high DM masses, these are nuclear recoils and multiphonon excitations, whereas for low masses, it is single-phonon generation, our focus here. Following the effective field theory of Refs. \cite{trickleMultichannelDirectDetection2020, trickleEffectiveFieldTheory2022}, the scattering rate per DM particle with velocity $\bm{v}$ is given by
\begin{equation}
    \Gamma(\boldsymbol{v})=\frac{\pi \bar{\sigma}_n}{\mu_{\chi n}^2} \int \frac{d^3 q}{(2 \pi)^3} \mathcal{F}_{\text {med }}^2(q) S\left(\boldsymbol{q}, \omega_{\boldsymbol{q}}\right).
\end{equation}
Here, we integrate over the momentum transfer $\bm{q} \equiv \bm{p} - \bm{p}'$, where $\bm{p} = m_\chi \bm{v}$ and $\bm{p}'$ are the momentum of the incident and scattered DM, respectively. The mediator form factor $\mathcal{F}_\text{med}(q)$ is given by

\begin{equation}
\mathcal{F}_{\text {med }}(q)= \begin{cases}1 & \text { (heavy mediator, $m_\phi \gg q_0$) } \\ \left(q_0 / q\right)^2 & \text { (light mediator, $m_\phi \ll q_0$) }\end{cases}
\end{equation}

We make the conventional choice for the reference momentum transfer $q_0 = m_\chi v_0$, with $v_0$ the DM's velocity dispersion \cite{trickleMultichannelDirectDetection2020}, and present our results in terms of a reference cross section $\bar{\sigma} \equiv \frac{\mu^2}{\pi} \overline{\left|\mathcal{M}(q_0)\right|}$, where $\mathcal{M}$ is the target-independent $\chi n \rightarrow \chi n$ vacuum matrix element (see SM).

The dynamic structure factor, $S(\bm{q}, \omega_{\bm{q}})$, encapsulates the target's response to the momentum transfer $\bm{q}$ and energy deposition $\omega_{\bm{q}} = \bm{q} \cdot \bm{v} - \frac{q^2}{2 m_\chi}$, and depends explicitly on the target's properties. For single-phonon excitations and the hadrophilic scalar mediator, we have \cite{trickleMultichannelDirectDetection2020} 

\begin{widetext}
\begin{equation}
\begin{aligned}
    & S\left(\bm{q}, \omega_{\bm{q}}\right)= 
    & \frac{\pi}{\Omega} \sum\limits_\nu \frac{1}{\omega_{\nu, \bm{k}}}\Bigg|\sum\limits_j \frac{A_j F_{N_j}(q)}{\sqrt{m_j}} e^{-W_j(\bm{q})} e^{i \bm{G} \cdot \bm{x}_j^0}\left(\bm{q} \cdot \boldsymbol{\epsilon}_{j,\nu,\bm{k}}^*\right)\Bigg|^2 \delta\left(\omega_{\bm{q}}-\omega_{\nu, \bm{k}}\right)
    \label{eq:dynStructFactor}
\end{aligned}
\end{equation}
\end{widetext}

The sum runs over all phonon branches $\nu$ and atoms $j$ in the primitive cell, with masses $m_j$, atomic mass numbers $A_j$, and equilibrium positions $\bm{x}_j^0$. $\Omega$ is the volume of the primitive cell, and  $\bm{k}$ is the phonon momentum in the first Brillouin zone (1BZ), constrainted to $\bm{k} = \bm{q} + \bm{G}$ by momentum conservation, for some reciprocal lattice vector $\bm{G}$. $\omega_{\nu,\bm{k}}$ and $\epsilon_{j,\nu,\bm{k}}$  are the phonon frequencies and polarization vectors of a given phonon mode $(\nu, \bm{k})$ and ion $j$. Finally, $F_{N_j}(q)$ and $W_j(\bm{q})$ are the nuclear form factor (see SM) and the Debye-Waller factor for atom $j$, respectively, with the latter given by

\begin{equation}
W_j(\bm{q})=\frac{\Omega}{4 m_j} \sum_\nu \int_{1 \mathrm{BZ}} \frac{d^3 k}{(2 \pi)^3} \frac{\left|\bm{q} \cdot \boldsymbol{\epsilon}_{j, \nu, \bm{k}}\right|^2}{\omega_{\nu, \bm{k}}}.
\label{eq:DWF}
\end{equation}
 Now, we can compute the total rate per unit target mass 
\begin{equation}
    R=\frac{1}{\rho_{\mathrm{T}}} \frac{\rho_\chi}{m_\chi} \int d^3 v f_\chi(\bm{v}) \Gamma(\bm{v}),
    \label{eq:total_rate}
\end{equation}
where $\rho_\chi = 0.3$ \GeV{}/cm$^3$ is the local DM energy density, $\rho_T$ is the target mass density, and $f_\chi$ is the incoming DM's velocity distribution in the target's rest frame (see SM). 

\begin{figure*}
    \includegraphics[width=\linewidth]{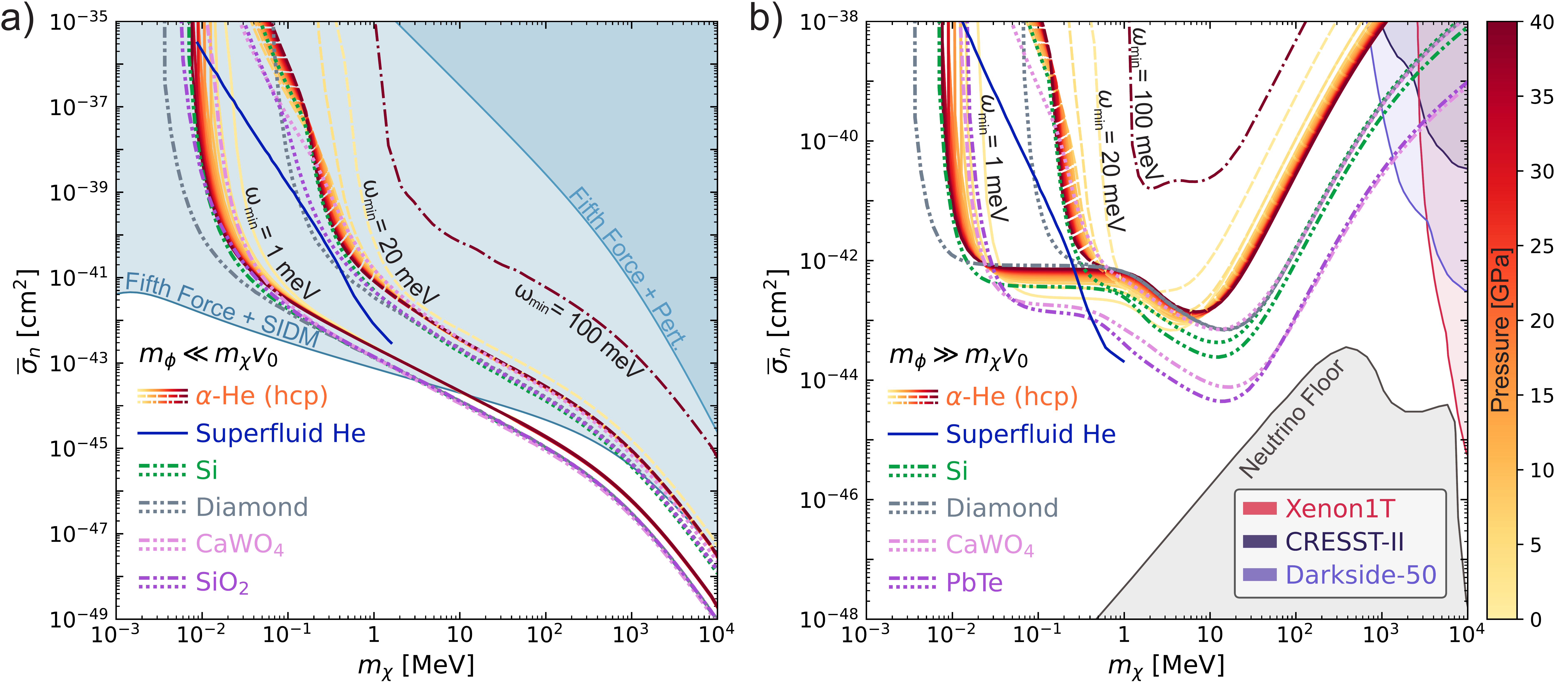}
    \caption{\label{fig:fig3} (a, b) Single-phonon projected reach for light and heavy scalar mediators, respectively. Gradient colored lines correspond to \hcpHe{} under different pressures between 1 (light orange) and 40 GPa (dark red), with solid colors representing other materials. For helium, solid, dashed, and dash-dotted lines correspond to 1 meV, 20 meV, and 100 meV detector thresholds ($\omega_\text{min}$), while for other materials, dash-double-dotted and dotted lines correspond to $\omegamin{} = 1$ and 20 meV. The solid navy line corresponds to multiphonon scattering in superfluid helium with $\omega_\text{min} \sim 1 \meV{}$ \cite{knapenLightDarkMatter2017}. For the light mediator, fifth force constraints are dominant \cite{knapenLightDarkMatter2017a}. DM self-interactions (SIDM) dominate if $m_\chi$ composes all the DM, but we only need perturbativity if it is a subcomponent. For the heavy mediator, we show nuclear recoil constraints from DarkSide50 \cite{DarkSide50}, Xenon1T \cite{Xenon1T.1,Xenon1T.2}, and CRESST-II \cite{CRESSTII}, as well as the neutrino floor \cite{NeutrinoFloor}.}
\end{figure*}

We used our package \texttt{DarkMAGIC} \cite{darkmagic}  for all calculations. The projected 95\% C.L. (3 events) exclusion reach on $\bar{\sigma}_n$, assuming zero background and a 100\% efficient detector, for an \hcpHe{} target with kg$\cdot$yr exposure is shown in Fig. \ref{fig:fig3}a (\ref{fig:fig3}b) for a light (heavy) scalar mediator, and pressures ranging from 1 GPa to 40 GPa. For comparison, we include the reach of the best-performing materials identified in Ref. \cite{griffinMultichannelDirectDetection2020}, as well as silicon and superfluid helium through multiphonon scattering \cite{knapenLightDarkMatter2017}. We consider three detector thresholds, $\omegamin{} = 1,\,20,\,100 \meV{}$. The same results are presented for \fccHe{} in Fig. \ref*{fig:fcc_He_reach} in the SM, for pressures ranging from 1 GPa to 12 GPa, the highest pressure at which this phase is stable. 

In the low-mass regime, both \hcpHe{} and \fccHe{} are competitive with silicon, even at the lower end of the pressure range. At much higher pressures near the maximum 40 GPa, \hcpHe{} is competitive with the best target identified in Ref. \cite{griffinMultichannelDirectDetection2020}, diamond, and both phases outperform silica (SiO$_2$) at the lower pressures. For both mediators, lower pressure targets lose reach as $\omegamin{}$ increases, and only the 40 GPa \hcpHe{} target has any reach at the highest threshold.%

For $\omegamin{} = 1 \meV{}$, higher pressure targets have better reach to lower $m_\chi$. For this benchmark model, the lowest accessible DM mass with acoustic phonons scales as $m_{\chi,\text{min}} \sim \omegamin{}/\csla{}$ \cite{griffinMultichannelDirectDetection2020}, and Fig. \ref{fig:fig2}d shows an increase in the speed of sound $\csla{}$ with pressure. For the light scalar mediator in Figs. \ref{fig:fig3}a and \ref*{fig:fcc_He_reach}a, the reach at high $m_\chi \gtrsim 0.1 \MeV$ is independent of target properties \cite{griffinMultichannelDirectDetection2020}, explaining the convergence of the curves at different pressures in that regime. 

We now focus on the heavy mediator, Figs. \ref{fig:fig3}b and \ref*{fig:fcc_He_reach}b. At $m_\chi \sim 20 \keV{}$, we find an inflection point in the $\omegamin{} = 1 \meV{}$ curves, beyond which low-pressure targets have better reach on $\bar{\sigma}_n$. Here, $q$ is still small enough to be inside the 1BZ, but is far from the $\Gamma$-point, and a parametric expansion of the rate shows a $1/\csla{}$ scaling \cite{griffinMultichannelDirectDetection2020}. At $m_\chi \sim  200 \keV{}$, the slope of the curves changes, but the ordering with pressure remains. Now $q$ is outside the 1BZ, leading to a rapidly oscillating $\bm{k}$ (and thus $\omega_{\nu\bm{k}}$), and the average phonon frequency in the 1BZ, $\omegaAvg$, is the relevant quantity. An analytic estimate of the rate shows $\omegaAvg{}^{-1}$ scaling \cite{griffinMultichannelDirectDetection2020}, and in Fig. \ref*{fig:freq_pres} we find that $\omegaAvg{}$ is an increasing function of pressure. In the final regime, with $m_\chi \gtrsim 8 \MeV$, $q$ becomes large enough that the Debye-Waller factor (Eq. \ref{eq:DWF}) is nonnegligible, cutting off the integral in Eq. \eqref{eq:total_rate}. The rate scales as $(A\omegaAvg)^2$, resulting in high-pressure targets having better reach. In these last three regimes, both helium phases are outperformed by targets with heavy atoms, such as PbTe and CaWO$_4$, which have low $\csla$ and $\omegaAvg$, and much larger mass numbers $A$. However, multiphonon scattering and nuclear recoil, which we do not include here, will dominate single-phonon scattering at large $m_\chi$. In all cases, single-phonon scattering in solid helium outperforms two-phonon scattering in superfluid helium due to the phase-space suppression in the latter. 

These trends generally hold at higher thresholds but with increasing contributions from optical phonons. In Fig. \ref{fig:fig3}b, we find an inflection point in the reach of \hcpHe{} at $\omegamin{} = 20 \meV{}$. Optical phonons have reach down to $m_{\chi, \text{min}} \sim \omega_\text{LO}$, scaling with the frequency of the longitudinal optical mode near the $\Gamma$ point, where the dispersion is flat. Conversely, $m_{\chi,\text{min}} \sim \omegamin{}/\csla{}$ for acoustic phonons. Consequently, the lowest $m_\chi$ regime is dominated by optical phonons at large $\omegamin{}$, leading to the inflection point. This feature is crucially missing from \fccHe{}, which lacks optical modes (see Fig. \ref*{fig:fcc_He_reach}). %

We also calculate the daily modulation rates for both phases as a function of pressure. Fig. \ref*{fig:daily_mod} shows a daily modulation in \hcpHe{} at the lowest accessible DM masses, ranging from 5\% (3\%) below and 10\% (5\%) above the average daily rate for a light (heavy) mediator. This modulation is approximately independent of pressure (see Fig. \ref*{fig:deltaR}). Conversely, the highly isotropic \fccHe{} exhibits no daily modulation.  However, since our DM rate is dependent on pressure (Figs.~\ref{fig:fig3} and \ref*{fig:fcc_He_reach}), it could potentially be used as a new method for background discrimination.

\inlinesect{Discussion} Our results show that pressure is an enticing route for novel tunable DM detector designs. We demonstrate this in highly compressible solid helium~\cite{grosseCompressibilitySolidNoble1964} up to 40 GPa but expect the trend to hold at higher pressures. While our findings are generally applicable to any solid, the impact of pressure is most pronounced in materials characterized by very high compressibility, offering an unexplored avenue for detector tunability and background discrimination with applied pressure. Nonetheless, we note that pressure generally reduces phonon lifetimes, usually scaling as $P^{-1}$ in, e.g., inorganic semiconductors \cite{ulrichLifetimePhononsSemiconductors1997,debernardiAnharmonicPhononLifetimes1995}. 

However, pressurized cryogenic detectors pose several engineering challenges. Older piston-cylinder-type cells are capable of reaching pressures beyond 6 GPa with relatively large sample volumes, $\mathcal{O}(\text{few cm}^3)$ \cite{gettingGasChargedPiston1994}. In contrast, Kawai-type multi-anvil presses \cite{ishiiBreakthroughPressureGeneration2019} can reach tens of gigapascals with sample volumes $\mathcal{O}(\text{few mm}^3)$. Diamond anvil cells (DACs), considered the workhorse of modern high-pressure physics, can operate at pressures beyond 400 GPa \cite{shenHighpressureStudiesXrays2016}.  DACs offer excellent optical access to the sample and can be used at cryogenic temperatures \cite{palmerSubKelvinMagneticElectrical2015}. However, the feasibility of incorporating a phonon sensor, such as a TES, into such apparatuses is uncertain. Nonetheless, the strategic design of a direct detection experiment incorporating a high-pressure cell, carefully balancing target size with achievable pressure, holds promise for optimizing the reach of compressible solid targets.

In the case of solid helium specifically, single crystals of the hcp phase can be grown from helium gas directly inside DACs \cite{zhaElasticityDensehelium2004,tateiwaEvaluationsPressuretransmittingMedia2009} or from the superfluid at cryogenic temperatures \cite{balibarGrowthDynamicshelium1991}, with some approaches yielding almost defect-free but fragile samples \cite{pantaleiHowPrepareIdeal2010a}. Further, it is possible to isotopically enrich helium-4 to less than 0.5 parts per trillion helium-3 \cite{hendryContinuousFlowApparatus1987}. Integrating these high-quality growth techniques with DACs could be advantageous for solid helium, as ultrahigh-quality crystals would significantly reduce scattering from dislocations and isotopic impurities, thereby improving phonon lifetimes. One potential approach involves using helium-3, characterized by a spin-1/2 nucleus, in a DAC with embedded nitrogen-vacancy (NV) centers. These NV centers have promising applications in quantum sensing \cite{hoRecentDevelopmentsQuantum2021,bhattacharyyaImagingMeissnerEffect2024}, and it might be possible to use them for phonon detection in materials with magnetic nuclei.

\inlinesect{Conclusion}
In conclusion, our findings highlight the effectiveness of hydrostatic pressure in enhancing the speed of sound and phonon frequencies in solid helium. This effect is particularly profound in highly compressible solids and elevates solid helium from a target without single-phonon reach to one that competes with leading targets like diamond, particularly at low DM masses. Despite the challenges involved in designing such experiments, exploring this additional degree of freedom is promising, as it not only introduces the possibility of using neglected targets but also offers the potential to fine-tune their properties to target different DM mass regimes and provide background discrimination.

\begin{acknowledgments}
We are grateful to S. Knapen for feedback on our manuscript and acknowledge helpful discussions with N. Taufertsh\"{o}fer, M. Garcia-Sciveres, and D. McKinsey. We thank K. Inzani and T. Trickle for sharing their reach data. This work was supported by the US Department of Energy under the Quantum Information Science Enabled Discovery (QuantISED) for High Energy Physics grant KA2401032. Work at the Molecular Foundry was supported by the Office of Science, Office of Basic Energy Sciences, of the U.S. Department of Energy under Contract No. DE-AC02-05CH11231. This research used resources of the National Energy Research Scientific Computing Center (NERSC), a Department of Energy Office of Science User Facility using NERSC award BES-ERCAP0028926.
\end{acknowledgments}

\bibliography{He,DarkMatter,Functionals,pseudos,HighPressure,Codes,Supp}

\end{document}